\documentclass[onecolumn,eqsecnum,aps,showkeys,showpacs,superscriptaddress, 11pt,prb]{revtex4-1}
\usepackage{epsfig}
\usepackage{dcolumn}
\usepackage{xspace}
\usepackage{color}
\usepackage[autostyle]{csquotes}
\usepackage{caption}
\usepackage{subcaption}

\begin{document}
\draft

\title{Collapse of ferromagnetism with Ti doping in Sm$_{0.55}$Sr$_{0.45}$MnO$_3$: A combined
experimental and theoretical study}

\author{P. Sarkar}
\email{psphysics1981@gmail.com}
\affiliation{Department of Physics,
Serampore College, Serampore 712201, India}
\author{N. Khan}
\email{nazirkhan91@gmail.com}
\author{K. Pradhan}
\email{kalpataru.pradhan@saha.ac.in}
\author{P. Mandal}
\email{prabhat.mandal@saha.ac.in} \affiliation{Saha Institute of
Nuclear Physics, HBNI, 1/AF Bidhannagar, Calcutta 700064, India.}

\date{\today}

\begin{abstract}
We have investigated the effect of Ti doping on the transport
properties coupled with the magnetic ones in
Sm$_{0.55}$Sr$_{0.45}$Mn$_{1-\eta}$Ti$_{\eta}$O$_3$ ($0 \leq \eta
\leq 0.04$). The parent compound, Sm$_{0.55}$Sr$_{0.45}$MnO$_3$, exhibits a first-order
paramagnetic-insulator to ferromagnetic-metal transition just below
$T_{\rm c}$ = 128 K. With substitution of Ti at Mn sites ($B$-site), $T_{\rm c}$
decreases approximately linearly at the rate of 22 K$\%^{-1}$ while
the width of thermal hysteresis in magnetization and resistivity
increases almost in an exponential fashion. The most spectacular
effect has been observed for the composition $\eta$=0.03, where a
magnetic field of only 1 T yields a huge magnetoresistance, $1.2
\times 10^7$ $\%$ at $T_c\approx$ 63 K. With increasing magnetic
field, the transition shifts towards higher temperature, and the
first-order nature of the transition gets weakened and eventually becomes
crossover above a critical field ($H_{cr}$) which increases with Ti
doping. For Ti doping above 0.03, the system remains insulting
without any ferromagnetic ordering down to 2 K. The Monte-Carlo
calculations based on a two-band double exchange model show
that the decrease of $T_{\rm c}$ with Ti doping is associated with the
increase of the lattice distortions around the doped Ti ions.

\vskip 1cm
\end{abstract}


 \maketitle

\section{INTRODUCTION}
Perovskite manganites of the form RE$_{1-x}$AE$_x$MnO$_3$ (RE:
rare-earth ions and AE: alkaline-earth ions) display rich varieties
of physical phenomena owing to complex interplay between spin,
charge, and orbital degrees of freedom.
\cite{rao,dago1,salamon,dago2,toku1,ons} The competition between
these degrees of freedom is most prominently manifested in
narrow-band system with large disorder. Usually, two types of
disorders are considered in manganites. One is $A$-site disorder,
namely the quenched disorder, that arises mainly due to the size
mismatch between RE and AE cations and the other is $B$(Mn)-site
disorder, originates due to the partial substitution of Mn by other
transition metal ions with different spin and valence state. Though,
$A$-site ions are not directly involve in charge conduction
mechanism, several studies have shown that the disorder at $A$-site has
a strong influence on different kinds of long-range ordering of
manganese sublattice. Among these  ordered phases, the
charge-ordered (CO) state is most sensitive to $A$-site disorder
while the ferromagnetic (FM)-metallic phase is relatively weakly
affected
\cite{hwang95,tokura96,rod,attfield,tomioka03,aka,goodenough,tomi,fisher,tomioka06,prb08,demko,prb09,prl09,sr17}.
On the other hand, the doping at $B$-site induces local disorder
directly into the Mn-O-Mn network and as a result, it has much
stronger effect on magnetic, transport and other physical properties
of the system as compared to $A$-site disorder. Only a few percent
of $B$-site doping can bring about a drastic change in the
electronic and magnetic  properties without a significant change in
the crystal structure. Several experiments have been performed on a
large number of combination of reference states and $B$-site dopants
\cite{reviewb,mori99,kimura,martin01,machida02,kimti,nam06,sakai07,nam08,lu09,ssmoru,dhiman,lusr}.
On the basis of reference state, two classes of materials can be
distinguished: (i) FM-metal at $x$$\approx$0.33$-$0.4 and (ii)
CO-insulator at around $x$$\approx$0.5\cite{pradhan3}. In half-doped CO manganites,
often substitution of small amount of Cr/Ni/Ru at Mn site
dramatically suppresses the long-range CO state and drives the
system into FM metallic state. On the contrary, the $B$-site doping
in FM manganites may result in a strong suppression of
ferromagnetism by localizing the charge carriers which lead to the
formation of inhomogeneous and insulating magnetic ground state. \\

The effect of Mn-site doping  on magnetic and transport properties
has already been studied extensively but mostly on wideband FM
manganites.\cite{reviewb,martin01,kimti,nam06,nam08,ssmoru,dhiman}
However, the role of Mn-site doping in narrowband FM system, in
particular, close to the multicritical point has not been studied in
details. In the present work, we focus on a narrowband manganite,
Sm$_{0.55}$Sr$_{0.45}$MnO$_3$ (SSMO), which locates near the
multicritical point where the three phases namely, FM-metal,
CO-insulator and antiferromagnetic-insulator compete strongly with
each other, to explore the role of Mn-site disorder on the FM phase
\cite{tomi,fisher,tomioka06,prb08,demko,prb09,prl09,sr17}. The
effect of $B$-site doping on magnetic and transport properties of
Sm$_{0.55}$Sr$_{0.45}$Mn$_{1-\eta}$Ti$_{\eta}$O$_3$ with $0 \leq
\eta \leq 0.04$ has been studied systematically. The results show
that FM-metal to paramagnetic (PM)-insulator transition in SSMO is
first-order with transition temperature, $T_c \approx$128 K. With
the substitution of non-magnetic Ti$^{4+}$, both the FM transition
temperature, $T_{\rm c}$, and metal-insulator transition temperature
(MIT), $T_{MI}$, decrease, while the thermal hysteresis width
($\Delta T$) in electrical resistivity ($\rho$) and magnetization
($M$) increases drastically. Only 3$\%$ Ti doping increases $\Delta
T$ from 4.5 to 23.4 K. To the best of our knowledge, such a huge
increase in $\Delta T$ due to the $B$-site substitution has not been
reported earlier in any FM manganite. The application of external
magnetic field ($H$) shifts MIT towards higher temperature, leading
to a field dependent phase boundary. Besides these experimental
findings, the role of $B$-site doping on transport and magnetic
properties has also been investigated using model Hamiltonian
calculations. Our calculations based on Monte-Carlo technique using a two-band double
exchange model including electron-phonon coupling, super-exchange
interactions and quenched disorder reveal that with increasing Ti content, the lattice distortions around the Ti ions
increases and as a result  $T_{\rm c}$ decreases, which qualitatively agree with experimental results.

\section{EXPERIMENTS}
Polycrystalline Sm$_{0.55}$Sr$_{0.45}$Mn$_{1-\eta}$Ti$_{\eta}$O$_3$
samples with $\eta$$=$0-0.04 were prepared by conventional
solid-state reaction technique. The starting materials, Sm$_2$O$_3$
(pre-fired), SrCO$_3$, Mn$_3$O$_4$ and TiO$_2$ were mixed in a
stoichiometric ratio and ground thoroughly in an agate mortar by
using ethanol. The mixture was put in a platinum crucible and
calcined in air at 1100$^\circ$C for few days with intermediate
grindings. The obtained powder was pulverized and sintered at
1200$^\circ$C for 24 h  to ensure the chemical homogeneity. Phase
purity and the structural analysis of the samples were done by
powder x-ray diffraction (XRD) technique with Cu-K$_{\alpha}$
radiation in a high resolution Rigaku x-ray diffractometer (TTRAX
II). For all the studied compositions ($\eta$$=$0-0.04), we did not
observe any peak due to the impurity phase in the XRD pattern. The
Rietveld refinement technique was used for structural analysis. The
dc magnetization measurements were performed using a  magnetic
property measurement system (SQUID-VSM, Quantum Design). Resistivity
measurements were performed by a conventional four-probe technique
over a wide range of temperature for different applied magnetic
fields up to 9 T. We have measured transport and magnetic properties
as functions of $H$ and $T$ for all the samples, but for clarity few
of them are presented.

\section{RESULTS AND DISCUSSION}

\subsection{Crystal Structure}

\begin{figure}[ht]
{\centering \resizebox{9cm}{13.5cm}{\includegraphics{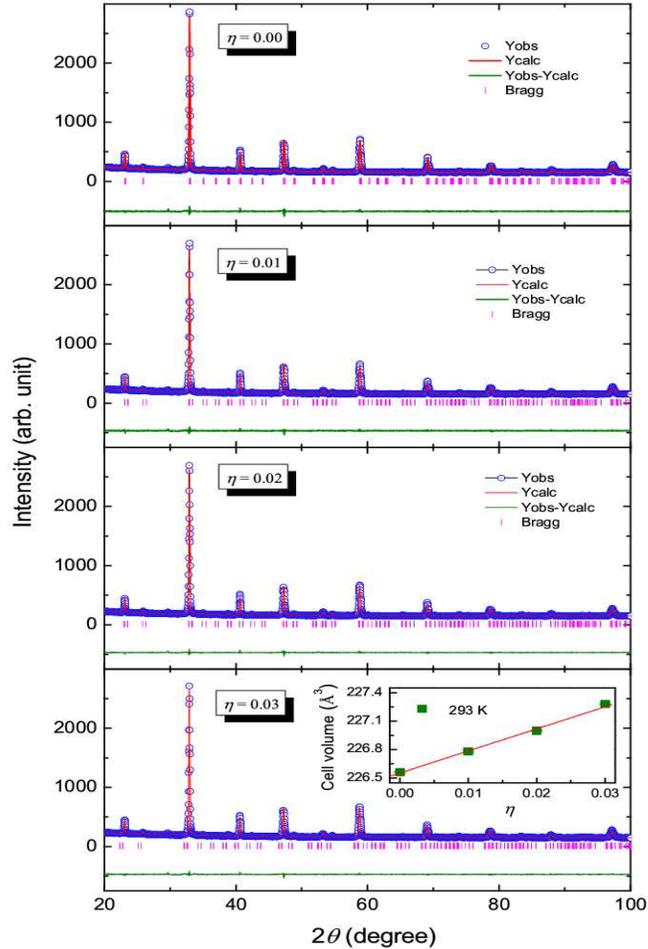}}\par}
\caption{(Color online) X-ray diffraction pattern  of
Sm$_{0.55}$Sr$_{0.45}$Mn$_{1-\eta}$Ti$_{\eta}$O$_3$ (0.00 $\leq$
$\eta$ $\leq$0.03) at room temperature. The bottom curves
(Yobs-Ycalc) are due to difference between the observed data and the
refinement data and the vertical bars indicate the Bragg peak
positions for Cu-$K_{\alpha_{1}}$ and Cu-$K_{\alpha_{2}}$
radiations. Inset shows Ti doping dependence of the unit cell volume
at room temperature.} \label{xrd.eps}
\end{figure}

Figure \ref{xrd.eps} shows the room-temperature powder x-ray
diffraction of Sm$_{0.55}$Sr$_{0.45}$Mn$_{1-\eta}$Ti$_{\eta}$O$_3$
for four compositions, $\eta$$=$0, 0.01, 0.02 and 0.03 as
representatives. The diffraction patterns show that all the samples
have a perovskite orthorhombic (space group $Pnma$) structure in
which the atomic positions of Sm(Sr): 4$c$($x$,1/4, \emph{z}),
Mn(Ti): 4$b$(0, 0, 1/2), O1: 4$c$($x$, 1/4, $z$) and O2: 8$d$($x$,
$y$, $z$) are used for indexing the Bragg peaks \cite{Murugesan}.
The crystal structure of the samples does not change with Ti doping.
However, with increasing Ti concentration, the lattice parameters
$a$, $b$, $c$ and hence the unit cell volume increase as shown in the inset of Fig. \ref{xrd.eps}. The increase
of cell volume suggests the substitution of Mn$^{4+}$ ions by
Ti$^{4+}$ ions, considering that the ionic radius of Ti$^{4+}$
(0.605 {\AA}) is larger than that of the Mn$^{4+}$ (0.530 {\AA}).
The refined parameters are presented in the Table I for various Ti
doping. Ti$^{4+}$ ions partially and randomly substitute isovalent
Mn$^{4+}$ ions and it is believed that like other Ti doped
manganites, the substitution of Ti in the present system also
increases the average (Mn, Ti)$-$O bond lengths, decreases
(Mn,Ti)$-$O$-$(Mn,Ti) bond angle and hence reduces the bandwidth of
the system \cite{kimti}.
\begin{table*}
{
\caption{Refined parameters for
Sm$_{0.55}$Sr$_{0.45}$Mn$_{1-\eta}$Ti$_{\eta}$O$_3$ (0.00$\le \eta
\le$0.03) compound at room-temperature with $Pnma$ space group.
\emph{a}, \emph{b}, \emph{c} are the lattice parameters and \emph{v}
is the unit cell volume. Numbers in the parenthesis are the
statistical errors. $U_{iso}$ is the isotropic atomic displacement
parameter. $\chi$$^2$ is goodness of the fit.}\label{I}
\begin{tabular*}{1.0\textwidth}{@{\extracolsep{\fill}}c c c c c}
\hline
\hline
$\rm Composition$ & $\eta$=0.00  & $\eta$=0.01 & $\eta$=0.02 & $\eta$=0.03 \\
\hline
\emph{a}({\AA}) & 5.43027(11) & 5.43271(13) & 5.43433(9) & 5.43703(13) \\

\emph{b}({\AA}) & 7.66747(14) & 7.66989(16) & 7.67356(11) & 7.67702(16) \\

\emph{c}({\AA}) & 5.44146(9) & 5.44253(11) & 5.44352(9) & 5.44512(12) \\

\emph{v}({\AA}$^3$) & 226.563(7) & 226.781(8) & 226.999(6) & 227.280(9) \\

$\chi$$^2$(\%)& 3.34 & 3.25 & 3.17 & 3.17 \\

$U_{iso}$({\AA}$^2$), Sm(Sr)& 0.0052(4) & 0.0067(4) & 0.0070(4)  & 0.0077(4)  \\

$U_{iso}$({\AA}$^2$), Mn(Ti)& 0.0021(6)  & 0.0049(6)  & 0.0032(5)  & 0.0048(6)  \\

$U_{iso}$({\AA}$^2$), O1(O2)& -0.003(2) & 0.007(2)  & 0.002(2) & 0.0033(19)  \\
\hline \hline
\end{tabular*}}
\end{table*}

\subsection{Magnetic and transport properties }
\begin{figure}[ht]
{\centering \resizebox{14cm}{10cm}{\includegraphics{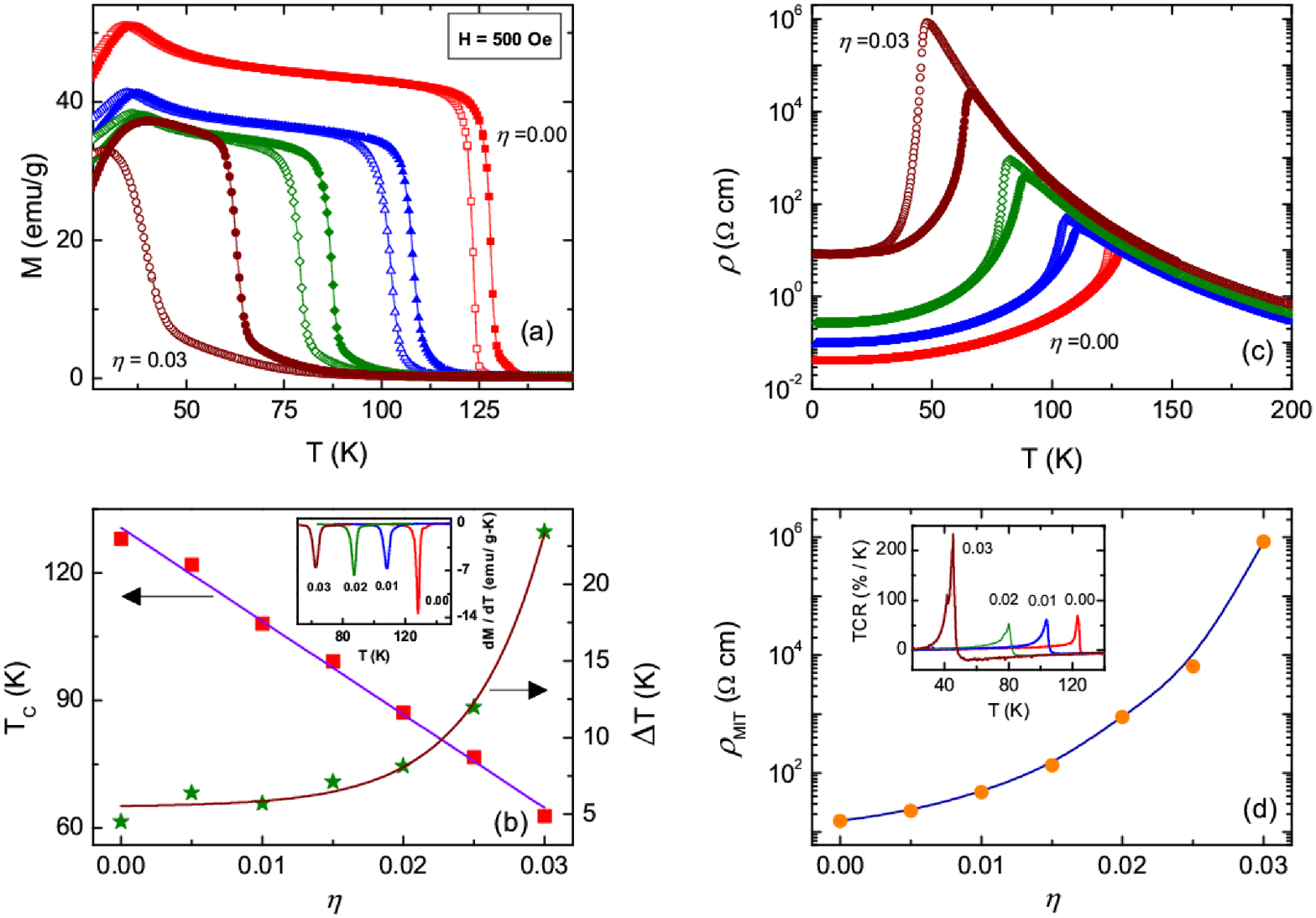}}\par}
\caption{(Color online) (a) Temperature ($T$) dependence of
magnetization ($M$) of
Sm$_{0.55}$Sr$_{0.45}$Mn$_{1-\eta}$Ti$_{\eta}$O$_3$ for $\eta$ =
0.0, 0.01, 0.02 and 0.03. Closed and open symbols represent heating
and cooling cycles, respectively. (b) Ferromagnetic-paramagnetic transition temperature, $T_{\rm c}$
(in the warming cycle) and thermal hysteresis width ($\Delta T$) as
a function of Ti concentration ($\eta$). Inset shows $T$ dependence
of d$M$/d$T$ for different $\eta$. (c) Temperature profile of
resistivity ($\rho$) for different $\eta$ both in warming (closed
symbol) and cooling (open symbol) cycles. (d) Peak resistivity at
metal insulator transition ($\rho_{MIT}$) as a function of $\eta$.
$\rho_{MIT}$ is derived from the cooling cycle of $\rho(T)$ curve.
Inset shows the temperature coefficient of resistivity (TCR) as a
function of $T$ for different $\eta$.} \label{mt.eps}
\end{figure}
Figure $\ref{mt.eps}$(a) shows the temperature dependence of
magnetization of Sm$_{0.55}$Sr$_{0.45}$Mn$_{1-\eta}$Ti$_{\eta}$O$_3$
for $\eta$$=$ 0.00, 0.01, 0.02 and 0.03. The parent compound, SSMO shows a sharp FM$-$PM transition at
$T_{\rm c}$$\sim$128 K, estimated as the temperature at which the
temperature coefficient of magnetization (d$M$/d$T$) exhibits a deep
minimum [inset of Fig. $\ref{mt.eps}$(b)]. However, the
magnetization data are not same in the warming and cooling cycles,
but exhibit a strong irreversibility of $\sim$4.5 K. The
irreversibility in $M$($T$) curve demonstrates the first-order
nature of FM transition in Sm$_{0.55}$Sr$_{0.45}$MnO$_3$. With
increasing $\eta$, the ferromagnetism is suppressed which is
indicated through the reduction of magnetization as well as a strong
decrease in $T_{\rm c}$. $T_{\rm c}$ is observed to decrease approximately
linearly with $\eta$ at the rate of 22 K$\%^{-1}$ [Fig.
$\ref{mt.eps}$(b)], which is much higher than that observed in
several other Ti doped FM manganites \cite{kimti,nam06,nam08}. Not
only the $T_{\rm c}$, the width of thermal hysteresis also changes
drastically with Ti doping which is shown in Fig. $\ref{mt.eps}$(b).
Remarkably, only 3$\%$ Ti doping increases $\Delta T$ from 4.5 K to
23.4 K.  It may be mentioned that $M$ has also been measured for
sample with slightly higher Ti concentration ($\eta$$=$0.04) but no
FM transition has been observed down to 2 K. In the resistivity
curve [Fig. $\ref{mt.eps}$ (c)], the MIT is observed at
$T_{MI}\approx$129 K ($\eta$$=$0), corresponding to resistivity
maximum.  The presence of thermal hysteresis in $\rho(T)$ curve
around $T_{MI}$ indicates that the MIT is first-order in nature.
Similar to magnetization, as Ti substitution proceeds, $T_{MI}$
decreases linearly while the width of the thermal hysteresis in
$\rho(T)$ increases exponentially. Depending on the degree of
influence of $B$-site doping on charge conduction, the whole
temperature region in $\rho(T)$ curve can be divided into three main
parts. At low temperatures well below $T_{MI}$, $\rho$ increases
sharply with increasing Ti content. As $\eta$ increases from 0 to
0.03, the residual resistivity increases almost by a factor 10$^4$.
The value of residual resistivity ($\sim$8.5 $\Omega$ cm) for
$\eta$$=$0.03 is well  above the Ioffe-Regel limit ($\sim10^{-3}$
$\Omega$ cm) to observe metallic behavior, suggesting that the
ground state is not a homogeneous ferromagnet rather it can be a
coexistence of FM and short-range CO states. Similar to residual
resistivity, the peak resistivity at MIT ($\rho_{MIT}$) also
enhances by a factor as high as 10$^5$ with increasing $\eta$ from 0
to 0.03, as shown in Fig. $\ref{mt.eps}$ (d). In the PM insulating
state well above $T_{MI}$, the effect of Ti substitution on $\rho$
is relatively weaker as compared to that in the low-temperature
region. For all the samples, the temperature coefficient of
resistivity [TCR = $\frac{1}{\rho}\left(\frac{d\rho}{dT}\right)$]
exhibits a very sharp peak at $T_{\rm c}$ [inset of Fig. $\ref{mt.eps}$
(d)], expected for a first-order phase transition. From figure, one
can see that the maximum value of TCR is almost same for 0$\leq \eta
\leq$0.02, but it abruptly increases for $\eta$ = 0.03. As MIT for
$\eta$$=$0.03 is much sharper as compared to other compositions, TCR
is very large for this composition in spite of large value of
$\rho$. This behavior is quite unexpected. Normally, disordering in
the active Mn-O-Mn network is supposed to broaden the FM transition.

\begin{figure}[ht]
{\centering \resizebox{14cm}{12cm}{\includegraphics{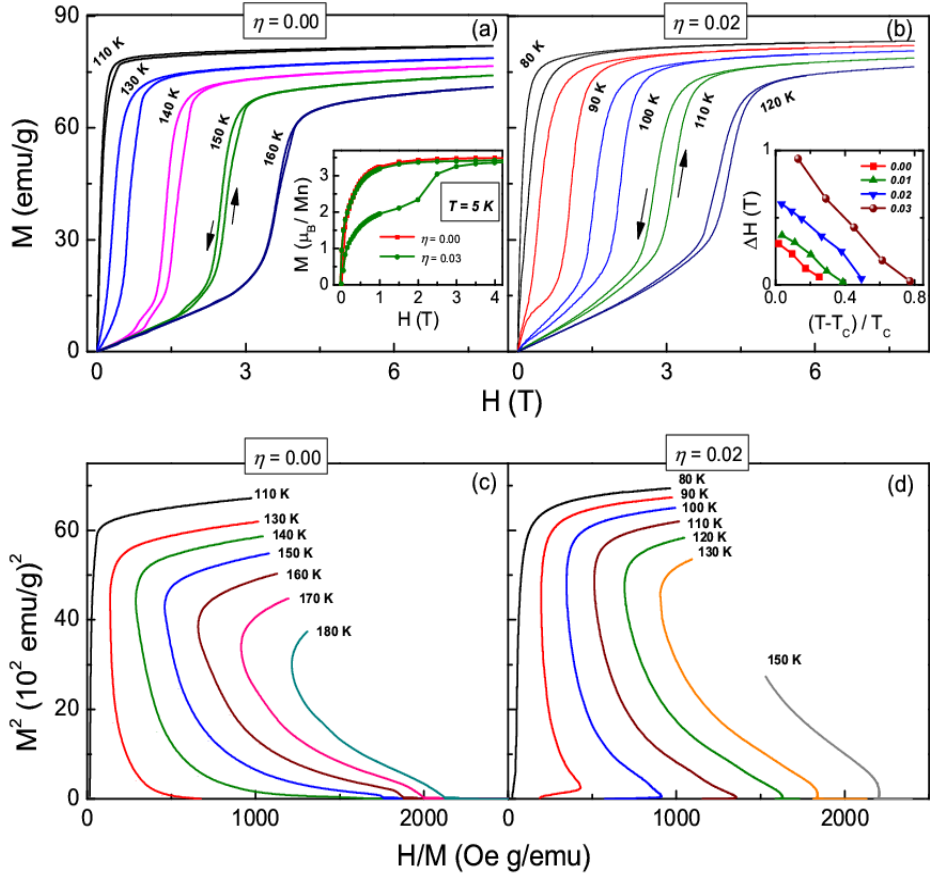}}\par}
\caption{(Color online) $M(H)$ isotherms of
Sm$_{0.55}$Sr$_{0.45}$Mn$_{1-\eta}$Ti$_{\eta}$O$_3$ for (a) $\eta$ =
0.00 and (b) $\eta$ = 0.02. Inset of figure (a) represents $M(H)$
hysteresis loop measured at 5 K, while the inset of figure (b) shows
hysteresis width ($\Delta H$) of $M(H)$ isotherms between increasing
and decreasing fields as a function of reduced temperature,
$(T-T_c)/T_c$. Arrott plots ($M^2$ vs. $H$/$M$) for (c) $\eta$ = 0.00
and (d) 0.02.} \label{mh.eps}
\end{figure}
The inset of Fig. $\ref{mh.eps}$ (a) shows magnetization hysteresis
loop at 5 K for $\eta$$=$0.00 and 0.03. We have measured $M(H)$ at 5 K
for $\eta$$=$0.00, 0.01, 0.02 and 0.03, but for clarity, only $\eta$$=$
0 and 0.03 data are presented in the figure. As in the case of a
typical soft ferromagnet, the magnetization of the samples with
$\eta \leq$0.02 increases rapidly with the application of field and
tends to saturate at a relatively low field strength. However, for
$\eta$$=$0.03, the nature of $M(H)$ curve at 5 K is not like a
simple ferromagnet but it exhibits a metamagnetic transition along
with field hysteresis. Ti substitution weakens the FM ordering of
parent compound and may favor the formation of short-range CO state.
The saturation magnetization for different $\eta$ are determined by
extrapolating the high field part of $M(H)$ curves to $H$$=$0. The
estimated values of saturation magnetization are 3.51, 3.46, 3.45
and 3.44 $\mu_B$ per Mn atom for $\eta$$=$0, 0.01, 0.02 and 0.03,
respectively. These values are slightly lower than their respective
theoretical values of  3.55, 3.52, 3.49 and 3.46 $\mu_B$ per Mn
atom, suggesting that the decrease of magnetization with Ti doping
is not only due the dilution of Mn$^{4+}$ atom but also due to the
weakening of exchange coupling. In the vicinity of  $T_{\rm c}$, $M(H)$
isotherms for $\eta$$=$0 and 0.02  are presented in Figs.
$\ref{mh.eps}$ (a) and (b), respectively. $M(H)$ curves below $T_{\rm c}$ are typical of
a ferromagnet with small hysteresis between increasing and
decreasing field. Initially, $M$ increases rapidly with $H$ and then
tends to saturate and the saturation value of $M$ gradually
decreases with increasing temperature. Above $T_{\rm c}$, we observe
S-shaped $M(H)$ isotherms, which indicates a metamagnetic phase
transition. With increasing $H$, first $M$ increases almost linearly
and then suffers a step-like jump, indicative of reentrant
ferromagnetism. Such a step-like jump in $M$ along with the
hysteresis are the manifestation of field-induced first-order PM-FM
phase transition. The inset of Fig. $\ref{mh.eps}$ (b) shows the
temperature [reduced temperature, $(T-T_c)/T_c$] dependence of width
of the field hysteresis ($\Delta H$) in $M(H)$ isotherms for different
$\eta$. For $\eta$$=$0, $\Delta H$ just above $T_{\rm c}$ ($\sim$128 K) is
$\sim$0.3 T, which decreases almost linearly with increasing $T$ and
eventually vanishes at a temperature that is around 1.3$T_{\rm c}$.
$\Delta H$ is observed to increase with increasing $\eta$ but
decreases  with increase in $T$ almost at the same rate as that for
$\eta$$=$0. Figures $\ref{mh.eps}$ (c) and (d) show the Arrott plots
($M^2$ vs. $H$/$M$), which offers a criterion for determining
whether FM to PM phase transition is first-order or second-order
purely by magnetic method \cite{arrott}. According to Banerjee
criterion, if the slope of the Arrott plot is positive then the FM
transition is second-order in nature and for a first-order
transition the slope is negative \cite{banerjee}. The undoped
compound shows a negative slope in $M^2$ vs. $H/M$ plot and this
behavior persists for $\eta \leq$ 0.03, which means that FM
transition in all samples is first-order.

\begin{figure}[ht]
{\centering \resizebox{14cm}{10cm}{\includegraphics{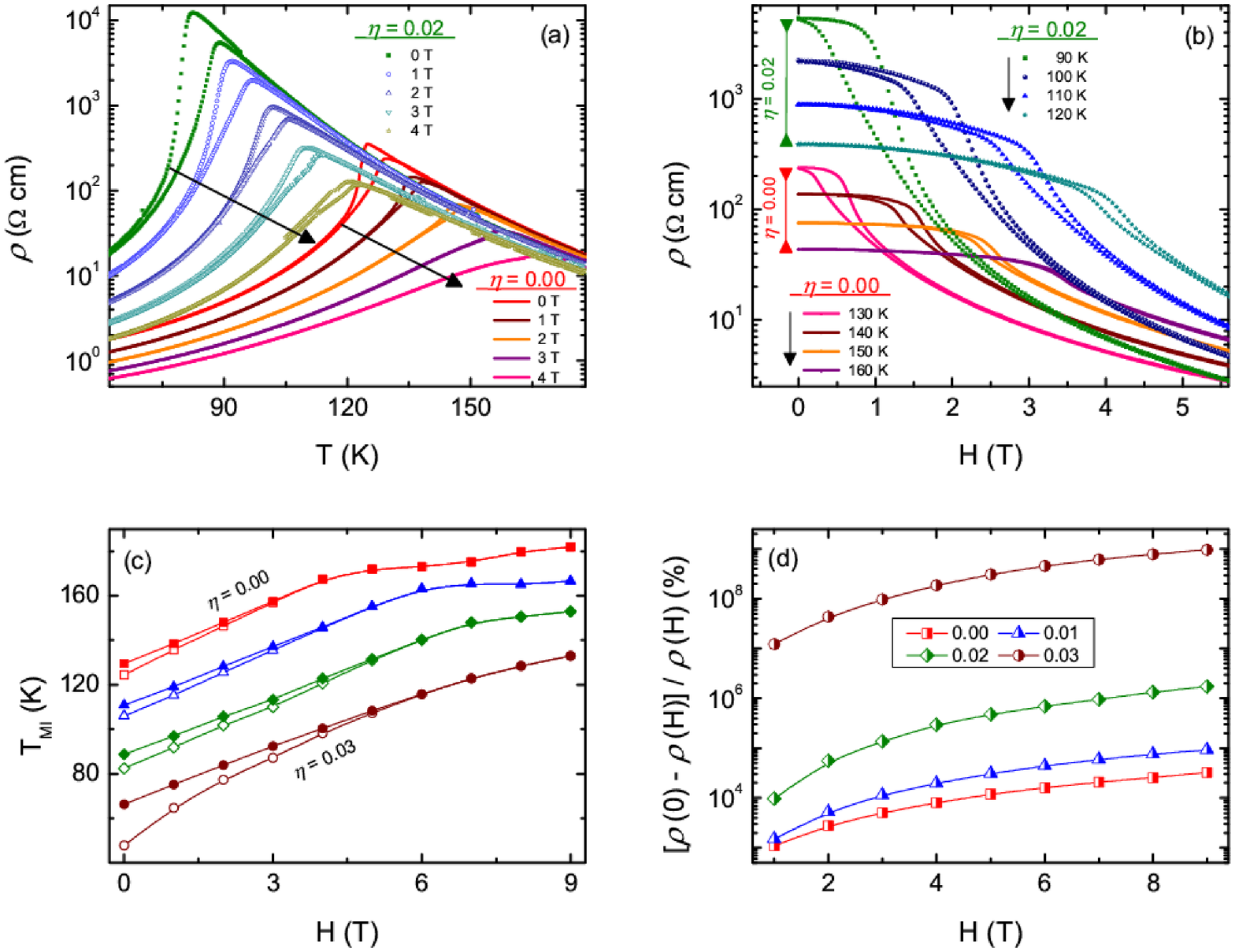}}\par}
\caption{(Color online) (a) $\rho$ vs $T$ curves at different $H$ (=
0, 1, 2, 3 and 4 T) for $\eta$ = 0.00 (lines) and 0.02 (symbols).
Data measured during warming and cooling cycles are shown
in the figure. The arrow indicates the direction of increasing
field. (b) $H$ dependence of $\rho$ at different temperatures for
$\eta$ = 0.00 (lines) and 0.02 (symbols). (c) The variation
of metal-insulator transition temperature ($T_{MI}$) with $H$ for
different $\eta$. Closed and open symbols are the $T_{MI}$'s derived
from the heating and cooling cycles of $\rho (T)$ curves,
respectively. (d) Magnetoresistance $\left[\left(\rho(0)-\rho
(H)\right)/\rho(H)\right]$ as a function of $H$ for different
$\eta$. It is calculated at $T = T_{MI}(H = 0)$.} \label{rt.eps}
\end{figure}
We now investigate the effect of external magnetic field on FM phase
transition in Sm$_{0.55}$Sr$_{0.45}$Mn$_{1-\eta}$Ti$_{\eta}$O$_3$
(0$\leq$$\eta$$\leq$0.03). The temperature and field dependence of
$\rho$ for different $\eta$ (0 and 0.02) are shown in Figs.
\ref{rt.eps} (a) and (b). Figure \ref{rt.eps} (a) shows that the
effect of $H$ on $\rho$ is maximum in the vicinity of $T_{MI}$,
whereas $\rho$ remains almost unchanged well above $T_{MI}$. For all
samples, the peak resistivity and resistivity below $T_{MI}$ are
observed to reduce strongly with field. Well below $T_{MI}$,
application of magnetic field enhances the spin-polarized tunneling
through grain boundaries and  as a result, residual resistivity
decreases rapidly with field. As shown in Fig. \ref{rt.eps} (b),
resistivity evolves with $H$ in an opposite way as that of
magnetization [Figs. \ref{mh.eps} (a) and (b)]. Just above $T_{\rm c}$,
$\rho$ drops sharply with $H$ along with field hysteresis, a
consequence of first-order phase transition. With increasing
temperature, the sharpness of the field-induced change in $\rho$
diminishes, the width of the field hysteresis gradually becomes
narrow and finally vanishes above a critical temperature, $T_{cr}$.
The resultant $T_{MI}-H$ phase diagram for various Ti concentrations
is plotted in Fig. \ref{rt.eps} (c). As $H$ increases, the width of
thermal hysteresis in $\rho$ gradually decreases and the two phase
transition lines, corresponding to the warming and cooling
processes, merge to one another at a critical magnetic field
($H_{cr}$). This feature indicates that external field suppresses
the first-order nature of the transition and the transition becomes
a crossover above $H_{cr}$, and the value of critical field $H_{cr}$
increases from 4 to 6 T as $\eta$  increases from o to 0.03. In the
regime of $H$$<$$H_{cr}$, $T_{MI}$ for all samples increases
linearly with $H$ at an average rate of 9 K/T but at a slower rate
above $H_{cr}$.  We have also calculated
magnetoresistance (MR) at $T$=$T_{MI}$ ($H$$=$0) for
0$\leq\eta\leq$0.03. Here, MR is defined as $MR$$=$$\Delta \rho/
\rho(H)$=$\left[\rho(0)-\rho(H)\right]/\rho(H) \times$100\%, where
$\rho(0)$ and $\rho(H)$ are the values of resistivity at zero field
and at an applied field $H$, respectively. Figure \ref{rt.eps} (d)
shows the typical magnetic field dependence of MR for different
$\eta$. For $\eta$$=$0, the value of MR at $H$$=$1 T is $1.1 \times
10^4$ $\%$, which increases with $H$ and becomes $3.2 \times 10^4$
$\%$ for $H$$=$9 T. MR enhances with Ti doping and the most
fascinating effect is observed for the composition $\eta$$=$0.03,
where MR reaches to $1.2 \times 10^7$ $\%$ for $H$$=$1 T only and it
becomes $\sim 10^9$ $\%$ for $H$$=$9 T. The observed value of MR is
much higher as compared with several other FM manganites.

\subsection{Theoretical Simulation}

We consider a two-band model Hamiltonian in two dimensions for
manganites in the large Hund's coupling limit ($J_H \rightarrow
\infty$) \cite{dago1,pradhan1} to study the role of Ti doping on transport
and magnetic properties of SSMO:
\begin{eqnarray}
\mathcal{H} &=& - \sum_{\langle ij \rangle \sigma}^{\alpha \beta}
{t}_{\alpha \beta}^{ij} d^{\dagger}_{i \alpha \sigma} d^{~}_{j \beta
\sigma} + J\sum_{\langle ij \rangle} {\bf S}_i.{\bf S}_j \cr &&  ~~
- \lambda \sum_i {\bf Q}_i.{\mbox {\boldmath $\tau$}}_i + {K \over
2} \sum_i {\bf Q}_i^2 + {\sum_i\epsilon_i n_i}, \nonumber
\end{eqnarray}
where $e_g$ electrons hop between nearest neighbor sites $i$ and $j$
with amplitude ${t}^{ij}_{\alpha \beta}$ (for two orbitals $a$ and
$b$). The hopping amplitudes ${t}^{ij}_{\alpha \beta}$ depend upon
the orientation of Mn $t_{2g}$ spins at the sites $i$ and $j$. For
details please see Ref.~\citenum{dago1}. $J$ and $\lambda$ are
anti-ferromagnetic super-exchange interactions between Mn
t$_{2g}$ spins (${\bf S}_{\rm i}$) and electron-phonon interactions
between the $e_g$ electrons and the Jahn-Teller phonons ${\bf Q}_i$
in the adiabatic limit, respectively. We treat all ${\bf S}_{\rm i}$
(with $|{\bf S}_i|=1$) and ${\bf Q}_i$ (with stiffness of the Jahn
Teller modes $K$=1) as classical~\cite{class-ref1,class-ref2}, and
measure all parameters ($J$, $\lambda$, and temperature $T$) in the
units hopping amplitude $t_{\rm aa}$.

\begin{figure}[ht]
{\centering
\resizebox{14.0cm}{10.0cm}{\includegraphics{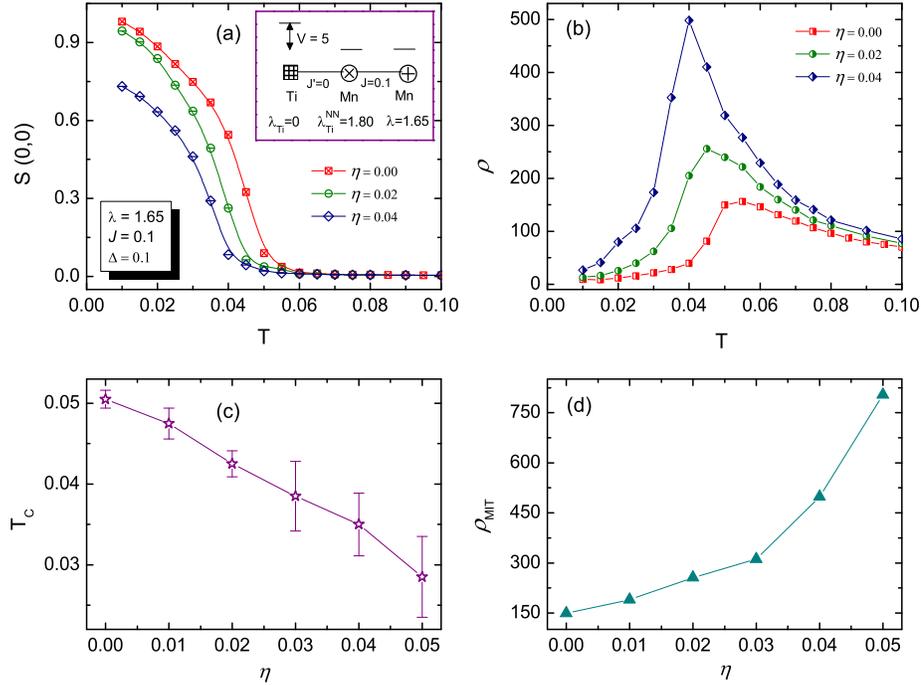}}\par}
\caption{(Color online) Temperature dependence of (a) ferromagnetic
structure factor [$S(0,0)$] and (b) resistivity ($\rho$) for dopant concentration $\eta$ = 0.00,
0.02, and 0.04. (c) Ferromagnetic transition temperature, $T_{\rm c}$, with various
dopant concentrations $\eta$. (d) Peak resistivity at metal insulator transition ($\rho_{\rm MIT}$) as a
function of $\eta$. Schematic figure in the inset of (a) shows the relevant levels on Ti and Mn
sites, and the coupling between these atoms. For notations please
see the text.} \label{th1.eps}
\end{figure}
This minimal model Hamiltonian $\mathcal{H}$ reproduces the correct
sequence of magnetic phases \cite{yunoki,pradhan2,pradhan3}. Typical
$\lambda$ $\sim$ 1.6$-$1.7 values with $J$ = 0.1 reproduce the
colossal magnetoresistive properties of intermediate bandwidth
manganites qualitatively \cite{pradhan2}. For $\lambda$ = 1.65 and
$J$ = 0.1, CE-type insulating phase can be reproduced for electron
density $n$ = 0.5, whereas the FM window spans over in the range
0.6$<n<$0.7 similar to intermediate bandwidth manganites. Here, we will
concentrate our calculations for $n = 0.65$ at which FM $T_{\rm c}$ is
maximum. Recall that in (Sm,Sr)
manganite system, $T_{\rm c}$ is optimum for $n$ = 0.55 \cite{tomioka06}
and our experiments are carried out at that electron density. The
effect of A-site disorder (due to mismatch between the ionic radii
of Sm$^{\rm 3+}$ and Sr$^{\rm 2+}$) is taken into account by adding
$\sum_i \epsilon_i n_i$ in the Hamiltonian where $\epsilon_i$ is the
quenched binary disorder potential with values $\pm \Delta$. We use
$\Delta$=0.1 and also checked our calculations for $\Delta$=0 and
0.2 to show that strong quenched disorder suppresses $T_{\rm c}$
more rapidly with Ti doping.

Next, in order to incorporate the effect of non-magnetic B-site dopants
(Ti$^{\rm 4+}$ in the present case), we modify the Hamiltonian as shown
schematically in the inset of Fig.~\ref{th1.eps} (a). A large energy
level $V$ (=5) is used at Ti sites, by adding $V \sum_i n_i$ to the
Hamiltonian \cite{pradhan1}. For $V$ = 5, the electron density
at the impurity site is close to zero. The electron-phonon coupling
is irrelevant at Ti sites, and for this reason we use $\lambda$ at impurity
sites $\lambda_{\rm Ti}$ = 0. Super-exchange interaction ($J'$)
between impurity site and nearest Mn-sites is modified to zero from $0.1$
[inset of Fig.~\ref{th1.eps} (a)]. Although our Hamiltonian has
spin moment at each impurity site, but that moment is very
weakly connected to rest of the system due to $J'=0$ and large on
site potential $V$ (= 5). So, these moments at impurity sites do not
affect the magnetism and are not taken into account while
calculating the ferromagnetic order. From our experimental results, it is
clear that with Ti doping, the unit cell volume of the system increases
(inset of Fig. \ref{xrd.eps}) and as a result, the bandwidth of the
system decreases \cite{kimti}. In order to take take this effect into
account, we modify the $\lambda$ values at the Mn
sites ($\lambda_{\rm Ti}^{\rm NN}$=1.80) those are nearest neighbor
to Ti ions. Recall that $\lambda$ is measured in units of kinetic energy,
and thus larger $\lambda$ corresponds to smaller bandwidth.
Large $\lambda_{\rm Ti}^{\rm NN}$ helps in localizing
the electrons at those sites and as a results Mn$^{\rm 3+}$ look-a-like
ions surrounds the Ti$^{\rm 4+}$ ions that minimizes the Coulomb
repulsion.

We use an exact diagonalization scheme to the itinerant electron
system for each configuration of the background classical variables
${\bf S}_{\rm i}$ and ${\bf Q}_{\rm i}$. We use a Monte Carlo
sampling technique based on the traveling cluster approximation \cite{tca-ref,pradhan2}
to access large system sizes. All physical quantities like
ferromagnetic structure factor and resistivity are thermally
averaged over ten different samples (starting from ten different
initial realizations of the quenched disorder and classical variables).

The temperature dependence of the ferromagnetic structure factor
$S$(0,0) and resistivity $\rho$ with Ti concentration $\eta$ is
shown in Figs.~\ref{th1.eps} (a) and (b) respectively. $S$(0,0) is
obtained by calculating $S(\textbf{q})$ = ${1 \over (N-\eta)^2}$
$\sum_{ij}$ $\bf {\bf S}_i\cdot {\bf S}_j$ e$^{i\bf{q}\cdot ({\bf
r}_i-{\bf r}_j)}$ at wave vector ${\bf q} = (0, 0)$ for Mn sites.
The resistivity, in units of ${\hbar a}/{\pi  e^2 }$ ($a$: lattice
constant), is obtained from the {\it dc} limit of the conductivity
(calculated using  the Kubo-Greenwood formalism)\cite{mahan-book,cond-ref}.
The $T_{\rm c}$ decreases approximately linearly and the resistivity
peak increases very fast with $\eta$ as shown in Figs.~\ref{th1.eps}
(c) and (d), respectively and agrees qualitatively with our
experiments.

\begin{figure}[ht]
{\centering
\resizebox{14.0cm}{6.5cm}{\includegraphics{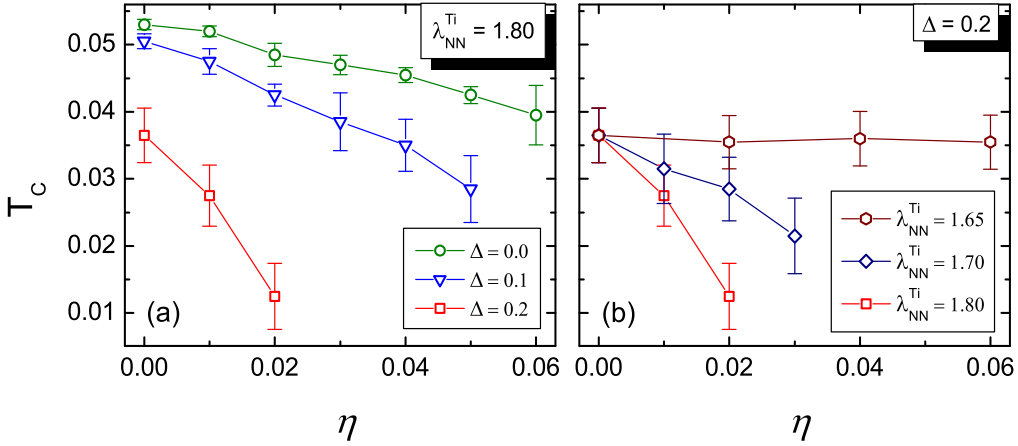}}\par}
\caption{(Color online) Variation of $T_{\rm c}$ with $\eta$ for
$\lambda$ = 1.65 and $J$ =0.1: (a) different $\Delta$ values (using
$\lambda_{\rm Ti}^{\rm NN}$=1.80) and (b) different $\lambda_{\rm Ti}^{\rm NN}$
values (using $\Delta$=0.2) } \label{th2.eps}
\end{figure}

For $\Delta=0.2$ ($\Delta=0$) and $\lambda_{\rm Ti}^{\rm NN}$ =
1.80, the FM $T_{\rm c}$ decreases faster (slower) than $\Delta=0.1$
case as shown in Fig. \ref{th2.eps} (a). This shows that disorder
also plays an important role in decreasing the FM $T_{\rm c}$. We
also use $\lambda_{\rm Ti}^{\rm NN}$ = 1.70 for $\Delta=0.2$ case
and find that the FM $T_{\rm c}$ decreases linearly, albeit up to
$\eta$=0.03, as shown in Fig. \ref{th2.eps} (b). But for
$\lambda_{\rm Ti}^{\rm NN}$ = 1.65 (\emph{i.e.}, without modifying
$\lambda$ at nearest neighbour Mn ions of Ti site) the FM $T_{\rm c}$ remains
more or less same until $\eta$=0.06 for $\Delta=0.2$. So, we believe
that lattice distortions around the Ti ions increase, which localize
the electrons and as a result the FM $T_{\rm c}$ decreases with Ti
doping.

\section{Conclusion}
The effect of Ti doping on the magnetotransport properties of
Sm$_{0.55}$Sr$_{0.45}$Mn$_{1-\eta}$Ti$_{\eta}$O$_3$ ($0 \leq \eta
\leq 0.03$) has been studied. All these samples undergo first-order
FM-metal to PM-insulator transition at $T_{\rm c}$ (or $T_{MI}$) along
with hysteresis. With increasing Ti concentration, $T_{\rm c}$ decreases
linearly while the magnetoresistance increases very rapidly. Our
theoretical calculations show that the FM $T_{\rm c}$ decreases due to the
increase of lattice distortion around the Ti ions. The application
of external field $H$ stabilizes the FM phase and thus weakens the
first-order nature of the transition. The critical magnetic field
where the first-order transition becomes a crossover increases with
Ti doping.

\end{document}